\documentclass[12pt]{article}

\begin{document}
\title{SPECTRA AND ELECTROMAGNETIC PROPERTIES OF THE NEUTRON-EXCESS
NUCLEUS $^{132}$In }
\author{V. I. Isakov\footnote{E-mail: Visakov@thd.pnpi.spb.ru}\\
\it Petersburg Nuclear Physics Institute, Gatchina\\
NRC Kurchatov Institute, Moscow, Russia}
\date{}
\maketitle

\noindent {\bf Abstract}

Spectra of levels and transition rates are calculated for the odd--odd particle--hole
neutron  excess nucleus  $^{132}$In in the framework of the RPA approach.\\

\bigskip
By now, nuclei far from the stability line including those close to the doubly magical ones
are of the great both experimental and theoretical interest. In our previous papers
\cite{Ero2001,Vadim16} we studied long chains of isotones having $N=82$ and $N=50$,
while in the papers \cite{Isakov2010,Isakov2013} we considered chains of isotopes with
$Z=28$ and $Z=50$, that include $^{78}$Ni, $^{100}$Sn and $^{132}$Sn. The $^{132}$Sn
and some neighboring nuclei were considered in the papers \cite{Ero1994,Ero1996}.
Odd--odd nuclei $^{132}$Sb and $^{134}$Sb, for which  experimental information
was available, were studied by us in details in  papers \cite{Mach1995,Isa161} and
\cite{Isakov2007,Isakov2012}. Recently, there appeared  experimental research \cite{Yungclaus2016},
where the authors determined  spectrum of low-lying levels in odd--odd  particle-hole
nucleus $^{132}$In, which is close to the doubly-magical $^{132}$Sn. Here, we carry out detailed
theoretical research of $^{132}$In including calculation of the spectra of levels and
electromagnetic characteristics of this nucleus.

The corresponding equations for determination of spectrum of levels may be obtained by
applying the algebra characteristic to the RPA approach, or by applying the Green functions
method. In the last case, the eigenvalues correspond to the poles of the particle-hole Green
function, where the particle-hole two-body matrix elements are used as an irreducible block
in the particle-hole channel (this method is nothing but the ladder approximation in the case
of particle--particle Green function).

For odd--odd particle-hole nuclei corresponding equations that define spectrum of levels and
amplitudes of states have the form
\begin{equation}
\left\| {A~~B \atop B~~C} \right\| \left({x \atop y}\right) =\ \omega_k \left({x \atop -y}\right),
\end{equation}
where
\begin{equation}
\left({x_{ab'} \atop y_{a'b}}\right)=f_{\alpha\beta}(\omega_k)=\langle JM(\omega_k)|
\sum_{m_\alpha m_\beta} C^{JM}_{j_\alpha m_\alpha j_\beta m_\beta}
(-1)^{\ell_\beta+j_\beta-m_\beta} a^+_{n_\alpha\ell_\alpha j_\alpha m_\alpha}
a_{n_\beta\ell_\beta j_\beta-m_\beta} |\tilde0\rangle.
\end{equation}
Here, dashed indices correspond to states below the Fermi surface, those without dashes -- to
states above the Fermi surface, while $|\tilde0\rangle$ corresponds
to the correlated ground state of the magical nucleus. In Eq.\,(1),
\begin{eqnarray}
&& A_{\alpha\beta,\mu\nu}=\,\langle j_\alpha\bar j_\beta J|\hat\vartheta|j_\mu\bar j_\nu J\rangle
+(\varepsilon_\alpha-\varepsilon_\beta) \delta_{\alpha\mu} \delta_{\beta\nu}; \quad
B_{\alpha\beta,\mu\nu} = \langle j_\alpha\bar j_\beta J|\hat\vartheta| j_\mu\bar j_\nu J\rangle;
\nonumber\\
 && C_{\alpha\beta,\mu\nu}\ =\ \langle j_\alpha\bar j_\beta J|\bar\vartheta| j_\mu\bar j_\nu J\rangle
 - (\varepsilon_\alpha-\varepsilon_\beta)\, \delta_{\alpha\mu} \delta_{\beta\nu}\,.
  \end{eqnarray}

 If we represent the two-body interaction $\hat\vartheta$ in the form
 $\hat\vartheta=\hat V^0+\hat V^1\,\vec\tau_1\vec\tau_2$, then we have for the particle-hole matrix
 elements entering formulas (3) the expression
 \begin{eqnarray}
 && \hspace*{-1cm}
 _a\langle\,j_\alpha\bar j_\beta J|\hat \vartheta|j_\mu\bar j_\nu J\,\rangle_a =\,-\sum_{J_0}
 (2J_0+1)W\Big[ j_\nu j_\mu j_\alpha j_\beta;JJ_0\Big]\ \times
 \\
 &&\times\ \bigg[\langle j_\nu j_\alpha J_0|\hat V^0-\hat V^1| j_\beta j_\mu J_0\rangle +
  (-1)^{j_\beta+j_\mu+J_0+1} \langle j_\nu j_\alpha J_0|2\hat V^1| j_\mu j_\beta J_0
  \rangle \bigg] (1)^{\ell_\beta+\ell_\nu}.
 \nonumber
 \end{eqnarray}

If we consider nucleus of the type ``magical $+p-n$'', then the indices ``$\alpha,\mu$'' correspond to protons $(\pi)$,
while ``$\beta,\nu$'' -- to neutrons $(\nu)$. In this case, the ``upper'' solutions $\omega^{+}_k$ of Eq.\,(1) correspond
to the nucleus ``magical$+p-n$'', while the ``lower'' ones $\omega^{-}_k$ -- to nucleus ``magical $-p+n$''.
The solutions of Eq.\,(1) are connected with the excitation energies by the relations
\begin{eqnarray}
&& E_k(p+ n^{-1})\ =\ \,\,\,\omega^{+}_k +B(Z+1,N-1)-B(Z,N)\,,
\nonumber\\
&& E_k(p^{-1}+n)\ =\ -\omega^{-}_k+B(Z-1,N+1)-B(Z,N)\,,
\end{eqnarray}
where $B$ are corresponding ground state binding energies.

For nucleus ``magical $+p-n$''  amplitudes ``$x$'' are large, while  amplitudes ``$y$'' are
small, and they appear only to the ground state correlations. At once, the situation is reverse if we
consider nucleus of the type ``magical$-p+n$'' $(\omega_{k}=\omega_{k}^{-})$.

Amplitudes ``$x$'' and ``$y$'' are normalized by the relation
\begin{equation}
\Big|\sum_{\alpha\beta} x^J_{j_\alpha j_\beta}(\omega_{k_1})x^J_{j_\alpha j_\beta}(\omega_{k_2})
- \sum_{\alpha\beta} y^J_{j_\alpha j_\beta}(\omega_{k_1}) y^J_{j_\alpha j_\beta}(\omega_{k_2})\Big|
\ =\ \delta_{k_1k_2}\,.
\end{equation}
If we define the reduced matrix element by the relation
\begin{equation}
\langle J'M'|\hat T_{\lambda_\mu}|JM\rangle\ =\ (-1)^{J'-M'} \left({J'~~\lambda~~J \atop -M'~~\mu~~M}
\right) \langle J'\|\hat T_\lambda\|J\rangle\,,
\end{equation}
then the reduced transition matrix element for the transition $|J(\omega^+_1)\rangle\to|J'(\omega^+_2)\rangle$
has the form
\begin{eqnarray}
&& \hspace*{-1cm}
\langle J'(\omega^+_2)\|\hat m(\lambda)\|J(\omega^+_1)\rangle\ =\ [(2J+1)(2J'+1)]^{1/2}\cdot \Big\{\sum_{\alpha\beta\mu}
\Big(x^J_{j_\alpha j_\beta}(\omega^+_1) x^{J'}_{j_\mu j_\beta}(\omega^+_2)\ -
\nonumber\\
&&-\ y^J_{j_\alpha j_\beta}(\omega^+_1)y^{J'}_{j_\mu j_\beta}(\omega^+_2)\Big) W[\lambda j_\mu Jj_\beta;
j_\alpha J'] \langle j_\mu\|\hat m(\lambda)\|j_\alpha\rangle\ \pm
\nonumber\\
&&\pm \sum_{\alpha\beta\nu}
\Big( x^J_{j_\alpha j_\beta}(\omega^+_1) x^{J'}_{j_\alpha j_\nu}(\omega^+_2)
- y^J_{j_\alpha j_\beta}(\omega^+_1) y^{J'}_{j_\alpha j_\nu}(\omega^+_2)\Big)\ \times
\nonumber\\
&& \times\  W[\lambda j_\beta J'j_\alpha;j_\nu J]\ \langle j_\nu \|\hat m(\lambda)\|j_\beta\rangle\Big\},
\end{eqnarray}
where the sign $(+)$ refers to $M\lambda$, while $(-)$  to $E\lambda$- transitions.
For nucleus ``magical $-p+n$'', when $\omega_{k} = \omega_{k}^-$, the expression (8) should be multiplied by $(-1)^\lambda$.

The quantities ``$\varepsilon$'' entering Eq. (3) are single-particle energies generated by the mean
field potential of the form,
\begin{equation}
U(\vec r,\vec\sigma)=U\cdot f(r)+U_{\ell s}\cdot \frac1r \frac{df}{dr}\,\vec\ell\vec s\,; \quad
f(r)=\frac1{1+\exp[(r-R)/a]}\,,
\end{equation}
where
$$
U=V_0\left(1-\beta\frac{N-Z}A t_Z\right), \quad U_{\ell s}=V_{\ell s}\left(1-\beta_{\ell s}
\frac{N-Z}At_Z\right), \quad R=r_0A^{1/3},
$$
$t_Z=1/2$, for neutrons and $t_Z=-1/2$ for protons. In the case of protons the Coulomb potential
of the uniformly charged sphere with radius $R_c=r_cA^{1/3}$ was added to (9).

The potential (9) was used by us in \cite{Isakov2002} and it ensures a good description of single-particle
spectra in nuclei near closed shells. In our calculations, we used the following values of the entering
parameters: $V_0=-51.55$\,MeV, $V_{\ell s}=32.4\rm\,MeV\cdot fm^2$, $a(p)=0.63\,$fm, $a(n)=0.66\,$fm, $\beta =1.31$,
$\beta_{\ell s}=-0.6$, $r_0=1.27\,$fm, $r_c=1.25\,$fm. For the better description of data on $^{132}$In, here in
our calculations we used experimental values of the single-particle energies, obtained from the data on the excitation and
binding energies  of the corresponding nuclei, if these data are available, see \cite{nds}--\cite{Hann2000}. These
single-particle energies are close to those generated by the potential (9).

The two-body interaction used by us here has the form
\begin{eqnarray}
\hat \vartheta & =& \hat V^0 +\hat V \vec\tau_1\vec\tau_2 =\ \Big(V+V_\sigma\vec\sigma_1\vec\sigma_2
  +V_T S_{12} +V_\tau\vec\tau_1\vec\tau_2 +V_{\tau\sigma}\vec\sigma_1\vec\sigma_2 \vec\tau_1\vec\tau_2
    +\ V_{\tau T}S_{12}\vec\tau_1\vec\tau_2\Big) \times
\nonumber\\
 &&\times\   \exp\Big(-\frac{r^2_{12}}{r^2_{00}}\Big)
  +\frac{e^2}{r_{12}}\Big(\frac12 -\hat t_Z(1)\Big) \Big(\frac12-\hat t_Z(2)\Big).
  \end{eqnarray}

In our calculations, we used two sets of parameters. The interaction RPA1 corresponds to the
following values of parameters: $V=-9.95$, $V_\sigma=2.88$, $V_T=-1.47$, $V_\tau=5.90$,
$V_{\tau\sigma}=4.91$, $V_{\tau T}=1.51$ (all these values are in MeV), and $r_{00}=1.8\,$fm.
These parameters were defined by us before \cite{Isakov1977}--\cite{Art89} from the
description of different experimental data near $^{208}$Pb and $^{146}$Gd.

In the RPA calculations of long chains of isotopes and isotones we used also the interaction RPA2
of the same form (10), but with the parameters $V=-16.65$, $V_\sigma=2.33$, $V_T=-3.00$, $V_\tau=3.35$,
$V_{\tau\sigma}=4.33$, $V_{\tau T}=3.00$ (MeV), and $r_{00}=1.75\,$fm. In the case of identical
particles this interaction coincides with that from the paper \cite{Heyde71}. The RPA2 interaction
also well reproduces the pattern of proton--neutron multiplet splitting in odd-odd nuclei close to
$^{208}$Pb.

Electromagnetic moments and transition rates were calculated by using the the effective multipole
operators of the form
\begin{eqnarray}
&& \hat m_{2\mu}(E2)\ =\ e^{p_1,n}_{\lambda=2}(eff)\cdot r^2 Y_{2\mu}(\theta,\varphi)\,,
\\
&& \hat m_{1\mu}(M1)=\mu_N\cdot\sqrt{\frac3{4\pi}} \left[g_\ell^{p,n}(eff)\cdot\vec\ell
+ g^{p,n}_s (eff)\cdot \vec s+g_2\cdot\tau_3\, r^2[Y_2\otimes\vec s]^1\right]_\mu\,.
\nonumber
\end{eqnarray}
Here, the values of gyromagnetic ratios and of the effective charges were the same as in our
previous papers \cite{Art82}, notably $e^p_{\lambda=2}(eff)=1.6|e|$, $e^n_{\lambda=2}(eff)=0.9|e|$,
$g^p_\ell(eff)=1.102$, $g^n_\ell(eff)=-0.005$, $g^p_s(eff)=3.79$, $g^n_s=-2.04$ and
$g_2=-0.031\rm\,fm^{-2}$. In Eq.\,(11) $\tau_3=+1$ for neutrons and $\tau_3=-1$ for protons.
\newpage

Results of our calculations are demonstrated in  Tables 1--6. Structure of states, as well as the
magnitudes of the magnetic dipole and electric quadrupole moments of the positive
and negative parity levels of $^{132}$In with excitation energies less than 2 MeV are shown in
Tables 1 and 2. Calculations demonstrate that most of states, especially those with the low
excitation energies,  are rather ``pure'', and may be characterized by the quantum numbers of the
leading configuration. The lowest levels are the states of negative parity and belong to the configuration
$\{\overline{\pi1g_{9/2}},\nu2f_{7/2}\}$. Energies of these levels as well as the character of the multiplet splitting
are in a good agreement with the results of the experiment, graphically shown in \cite{Yungclaus2016}.
In Tables 3 and 4 we show  calculated values of the $E2$ transition rates $B(E2)$ between the
states of the positive (Table 3) and negative (Table 4) parities. Note, that in case of the lowest
levels that belong to the configuration $\{\overline{\pi1g_{9/2}},\nu2f_{7/2};J\}$, transitions with
$\Delta J=1$ are enhanced, while the transitions with $\Delta J=2$ are retarded. This is opposite
as compared to levels of the lowest proton-neutron multiplet $\{\pi1g_{7/2},\nu2f_{7/2};J\}$ in
$^{134}$Sb, where the transitions with $\Delta J=1$ are retarded, while the transitions with
$\Delta J=2$ are enhanced \cite{Isakov2012}. In Tables 5 and 6 one can see our results relating to
the $M1$ transition rates. Parallel with the $B(M1)$ values, we also show here signs of the
ratios $\langle J^{\pi}_f||\hat{m}(E2)||J^{\pi}_i \rangle/\langle J^{\pi}_f||\hat{m}(M1)||J^{\pi}_i\rangle$,
that are obtained in our calculations. Together with other data from  Tables 3--6 this enables one to
calculate the values of the mixing parameter $\delta$, that may be defined in the angular correlation
experiments.

\vspace{0.5cm}
This work was supported by the RSF grant No 14--22--00281.

\begin{table} 
\caption{Energies (in MeV) and electromagnetic moments of the positive
      parity levels in $^{132}$In. Magnetic moments are in the units of
      $\mu_N$, while the quadrupole moments are in the units  \newline
      of $|e|\cdot\rm fm^2$. Electromagnetic moments are calculated by
      using the interaction RPA2.}

\begin{center}
\begin{tabular}{||c|c|c|c|c|c||}
\hline\hline
     Level & Energy &  Energy &  Magnetic &  Quadrupole   &   Leading \\
           &   RPA1   &  RPA2  & moment   & moment   &  configuration \\
\hline\hline
     $0^+_1$  & 1.852  & 1.799 & 0.000  &  0.000 & $\overline{\pi2p_{1/2}},\nu3p_{1/2}$ \\
     $1^+_1$  & 1.511  & 1.563 & --0.110 E+01 &  --0.774 E+01 & $\overline{\pi2p_{1/2}},\nu3p_{3/2}$\\
     $1^+_2$  & 1.954  & 1.982 &  0.419 E+00  &   0.958  E+00 & $\overline{\pi2p_{1/2}},\nu3p_{1/2}$\\
     $2^+_1$  & 1.474  & 1.509 & --0.838 E+00 &  --0.157 E+02 & $\overline{\pi2p_{1/2}},\nu3p_{3/2}$\\
     $2^+_2$  & 2.083  & 2.084 & --0.254 E+01 &  --0.910 E+01 & $\overline{\pi2p_{3/2}},\nu2f_{7/2}$\\
     $3^+_1$  & 0.488  & 0.507 & --0.137 E+01 &  --0.216 E+02 & $\overline{\pi2p_{1/2}},\nu2f_{7/2}$\\
     $3^+_2$  & 1.848  & 1.871 & --0.586 E+00 &  --0.213 E+02 & $\overline{\pi2p_{3/2}},\nu2f_{7/2}$\\
     $4^+_1$  & 0.510  & 0.533 & --0.819 E+00 &  --0.251 E+02 & $\overline{\pi2p_{1/2}},\nu2f_{7/2}$\\
     $4^+_2$  & 1.743  & 1.746 &   0.315 E+00 &  --0.162 E+02 & $\overline{\pi2p_{3/2}},\nu2f_{7/2}$\\
     $4^+_3$  & 2.036  & 2.025 &   0.484 E+00 &  --0.273 E+02 & $\overline{\pi2p_{1/2}},\nu1h_{9/2}$\\
     $5^+_1$  & 1.878  & 1.912 &   0.178 E+01 &  --0.992 E+01 & $\overline{\pi2p_{3/2}},\nu2f_{7/2}$ \\
     $5^+_2$  & 1.961  & 1.972 &   0.128 E+01 &  --0.240 E+02 & $\overline{\pi2p_{1/2}},\nu1h_{9/2}$\\
  \hline\hline

  \end{tabular}
  \end{center}
  \end{table}

\begin{table} 
\caption{Energies (in MeV) and electromagnetic moments of the negative
 parity levels in $^{132}$In. Magnetic moments are in the units of $\mu_N$,
 while the quadrupole moments are in the units \newline of $|e|\cdot\rm fm^2$.
 Electromagnetic moments are calculated by using the interaction RPA2.}

\begin{center}
\begin{tabular}{||c|c|c|c|c|c||}
\hline\hline
   Level & Energy & Energy &  Magnetic  &   Quadrupole  &     Leading  \\
         & RPA1   & RPA2   &  moment &   moment  &     configuration \\
\hline\hline
$1^-_1$ & 0.926 & 0.821 &  0.443 E+01 & 0.408 E+01 & $\overline{\pi1g_{9/2}},\nu2f_{7/2}$\\
$2^-_1$ & 0.356 & 0.381 &  0.389 E+01 & 0.944 E+01 & $\overline{\pi1g_{9/2}},\nu2f_{7/2}$\\
$2^-_2$ & 2.101 & 2.117 &  0.176 E+01 & 0.813 E+00 & $\overline{\pi1g_{9/2}},\nu1h_{9/2}$\\
$3^-_1$ & 0.171 & 0.167 &  0.368 E+01 & 0.153 E+02 & $\overline{\pi1g_{9/2}},\nu2f_{7/2}$\\
$3^-_2$ & 1.169 & 1.164 &  0.641 E+01 & 0.174 E+02 & $\overline{\pi1g_{9/2}},\nu3p_{3/2}$\\
$3^-_3$ & 1.931 & 1.910 &  0.252 E+01 & 0.116 E+01 & $\overline{\pi1g_{9/2}},\nu1h_{9/2}$\\
$4^-_1$ & 0.105 & 0.113 &  0.384 E+01 & 0.167 E+02 & $\overline{\pi1g_{9/2}},\nu2f_{7/2}$\\
$4^-_2$ & 0.884 & 0.891 &  0.628 E+01 & 0.357 E+02 & $\overline{\pi1g_{9/2}},\nu3p_{3/2}$\\
$4^-_3$ & 1.472 & 1.489 &  0.545 E+01 & 0.317 E+02 & $\overline{\pi1g_{9/2}},\nu3p_{1/2}$\\
$4^-_4$ & 1.768 & 1.805 &  0.330 E+01 & 0.180 E+01 & $\overline{\pi1g_{9/2}},\nu1h_{9/2}$\\
$4^-_5$ & 2.127 & 2.156 &  0.507 E+01 & 0.146 E+02 & $\overline{\pi1g_{9/2}},\nu2f_{5/2}$\\
$5^-_1$ & 0.040 & 0.041 &  0.404 E+01 & 0.153 E+02 & $\overline{\pi1g_{9/2}},\nu2f_{7/2}$\\
$5^-_2$ & 0.834 & 0.831 &  0.521 E+01 & 0.360 E+02 & $\overline{\pi1g_{9/2}},\nu3p_{3/2}$\\
$5^-_3$ & 1.474 & 1.483 &  0.695 E+01 & 0.334 E+02 & $\overline{\pi1g_{9/2}},\nu3p_{1/2}$\\
$5^-_4$ & 1.739 & 1.737 &  0.409 E+01 & 0.171 E+01 & $\overline{\pi1g_{9/2}},\nu1h_{9/2}$\\
$5^-_5$ & 2.131 & 2.139 &  0.596 E+01 & 0.161 E+02 & $\overline{\pi1g_{9/2}},\nu2f_{5/2}$\\
$6^-_1$ & 0.049 & 0.059 &  0.445 E+01 & 0.156 E+02 & $\overline{\pi1g_{9/2}},\nu2f_{7/2}$\\
$6^-_2$ & 0.993 & 1.022 &  0.532 E+01 & 0.190 E+02 & $\overline{\pi1g_{9/2}},\nu3p_{3/2}$\\
$6^-_3$ & 1.658 & 1.692 &  0.489 E+01 & 0.365 E+01 & $\overline{\pi1g_{9/2}},\nu1h_{9/2}$\\
$6^-_4$ & 2.023 & 2.041 &  0.640 E+01 & 0.180 E+02 & $\overline{\pi1g_{9/2}},\nu2f_{5/2}$\\
$7^-_1$ & gr.st.& gr.st &  0.478 E+01 & 0.133 E+02 & $\overline{\pi1g_{9/2}},\nu2f_{7/2}$\\
$7^-_2$ & 1.695 & 1.708 &  0.570 E+01 & 0.581 E+01 & $\overline{\pi1g_{9/2}},\nu1h_{9/2}$\\
$8^-_1$ & 0.210 & 0.253 &  0.527 E+01 & 0.130 E+02 & $\overline{\pi1g_{9/2}},\nu2f_{7/2}$\\
$8^-_2$ & 1.620 & 1.649 &  0.648 E+01 & 0.719 E+01 & $\overline{\pi1g_{9/2}},\nu1h_{9/2}$\\
$9^-_1$ & 1.997 & 2.059 &  0.729 E+01 & 0.110 E+02 & $\overline{\pi1g_{9/2}},\nu1h_{9/2}$\\
\hline\hline

\end{tabular}
\end{center}
\end{table}


  \newpage

  \begin{table} 
\caption{ $E2$ transition rates $B(E2)$ in units of $e^2\cdot\rm fm^4$
      between the levels of positive parity \newline in $^{132}$In.
      Calculations are performed by using the interaction RPA2.}
      \begin{center}
      \begin{tabular}{||c|c||c|c||}
      \hline\hline

     Transition $i \to f $ &   $B(E2; i \to f)$  &   Transition $i \to f$  &  $B(E2; i \to f)$\\

     \hline \hline

   $2^+_1 \to 0^+_1$ &  0.454 E+02 &  $3^+_2 \to 3^+_1$ &  0.264  E+02\\

   $2^+_2 \to 0^+_1$  &  0.837 E+01 & $4^+_1 \to 3^+_1$ &  0.813  E+01\\

   $1^+_2 \to 1^+_1$  &  0.889 E+02 & $4^+_2 \to 3^+_1$  &  0.744 E+02\\

   $2^+_1 \to 1^+_1$  &  0.243 E+02 & $4^+_3 \to 3^+_1$  &  0.303 E+01\\

   $2^+_2 \to 1^+_1$  &  0.287 E--01 & $5^+_1 \to 3^+_1$  &  0.506 E+02\\

   $3^+_1 \to 1^+_1$  &  0.801 E+02 & $5^+_2 \to 3^+_1$  &  0.140 E+02\\

   $3^+_2 \to 1^+_1$  &  0.161 E+01 & $4^+_1 \to 3^+_2$  &  0.488 E+02\\

   $2^+_1 \to 1^+_2$  &  0.601 E+02 & $4^+_2 \to 3^+_2$  &  0.558 E+02\\

   $2^+_2 \to 1^+_2$  &  0.231 E+01 & $4^+_3 \to 3^+_2$  &  0.737 E--02\\

   $3^+_1 \to 1^+_2$  &  0.949 E--02 & $5^+_1 \to 3^+_2$  &  0.263 E+02\\

   $3^+_2 \to 1^+_2$  &  0.121 E+01 & $5^+_2 \to 3^+_2$  &  0.880 E+01\\

   $2^+_2 \to 2^+_1$  &  0.203 E+00 & $4^+_2 \to 4^+_1$  &  0.537 E+02\\

   $3^+_1 \to 2^+_1$  &  0.124 E+02 & $4^+_3 \to 4^+_1$  &  0.241 E--01\\

   $3^+_2 \to 2^+_1$  &  0.280 E+01 & $5^+_1 \to 4^+_1$  &  0.544 E+02\\

   $4^+_1 \to 2^+_1$  &  0.972 E+02 & $5^+_2 \to 4^+_1$  &  0.229 E+01\\

   $4^+_2 \to 2^+_1$  &  0.891 E+00 & $4^+_3 \to 4^+_2$  &  0.618 E--01\\

   $4^+_3 \to 2^+_1$  &  0.149 E--02 & $5^+_1 \to 4^+_2$  &  0.845 E+02\\

   $3^+_1 \to 2^+_2$  &  0.170 E+02 & $5^+_2 \to 4^+_2$  &  0.203 E+02\\

   $3^+_2 \to 2^+_2$  &  0.902 E+00 & $5^+_1 \to 4^+_3$  &  0.215 E+01\\

   $4^+_1 \to 2^+_2$  &  0.309 E+02 & $5^+_2 \to 4^+_3$  &  0.673 E+01\\

   $4^+_2 \to 2^+_2$  &  0.380 E+02 & $5^+_2 \to 5^+_1$  &  0.100 E+02\\

   $4^+_3 \to 2^+_2$  &  0.606 E+01  &&\\

 \hline\hline
 \end{tabular}
 \end{center}
 \end{table}

 \begin{table} 

\caption{
 $E2$ transition rates $B(E2)$ in units of $e^2\cdot\rm fm^4$
     between the states of negative parity \newline in $^{132}$In.
     Calculations are performed by using the interaction RPA2.}

\begin{center}

\begin{tabular}{||c|c||c|c||} \hline\hline

        Transition $i \to f$   &    $B(E2; i \to f)$  & Transition $i \to f$   &    $B(E2; i \to f)$\\
        \hline\hline

      $2^-_1 \to 1^-_1$    &   0.147 E+02 &  $6^-_2 \to 4^-_1$    &   0.145 E+00\\

      $3^-_1 \to 1^-_1$    &   0.623 E+00 &  $5^-_1 \to 4^-_2$    &   0.296 E+02\\

      $3^-_2 \to 1^-_1$    &   0.290 E+02 &  $5^-_2 \to 4^-_2$    &   0.567 E+02\\

      $3^-_1 \to 2^-_1$    &   0.398 E+02 &  $6^-_1 \to 4^-_2$    &   0.176 E+02\\

      $3^-_2 \to 2^-_1$    &   0.576 E+02 &  $6^-_2 \to 4^-_2$    &   0.520 E+01\\

      $4^-_1 \to 2^-_1$    &   0.760 E--01 &  $5^-_2 \to 5^-_1$    &   0.344 E+02\\

      $4^-_2 \to 2^-_1$    &   0.778 E+01 &  $6^-_1 \to 5^-_1$    &   0.189 E+03\\

      $3^-_2 \to 3^-_1$    &   0.466 E+02 &  $6^-_2 \to 5^-_1$    &   0.997 E+01\\

      $4^-_1 \to 3^-_1$    &   0.994 E+02 &  $7^-_1 \to 5^-_1$    &   0.243 E+00\\

      $4^-_2 \to 3^-_1$    &   0.377 E+02 &  $6^-_1 \to 5^-_2$    &   0.492 E+02\\

      $5^-_1 \to 3^-_1$    &   0.936 E--01 &  $6^-_2 \to 5^-_2$    &   0.109 E+03\\

      $5^-_2 \to 3^-_1$    &   0.963 E+00 &  $7^-_1 \to 5^-_2$    &   0.432 E+02\\

      $4^-_1 \to 3^-_2$    &   0.141 E+02 &  $6^-_2 \to 6^-_1$    &   0.157 E+02\\

      $4^-_2 \to 3^-_2$    &   0.129 E+01 &  $7^-_1 \to 6^-_1$    &   0.199 E+03\\

      $5^-_1 \to 3^-_2$    &   0.455 E+01 &  $8^-_1 \to 6^-_1$    &   0.168 E+01\\

      $5^-_2 \to 3^-_2$    &   0.116 E+02 &  $7^-_1 \to 6^-_2$    &   0.154 E+02\\

      $4^-_2 \to 4^-_1$    &   0.550 E+02 &  $8^-_1 \to 6^-_2$    &   0.744 E+02\\

      $5^-_1 \to 4^-_1$    &   0.162 E+03 &  $8^-_1 \to 7^-_1$    &   0.131 E+03\\

      $5^-_2 \to 4^-_1$    &   0.124 E+02 &  $9^-_1 \to 7^-_1$    &   0.226 E+00\\

      $6^-_1 \to 4^-_1$    &   0.267 E+00 &  $9^-_1 \to 8^-_1$    &   0.370 E+01 \\

       \hline\hline
       \end{tabular}  \end{center}

       \end{table}

   \newpage

  \begin{table}  
  \caption{
 $M1$ transition rates $B(M1)$ between the levels of positive parity
 in $^{132}$In in units of $\mu_N^2$. Signs in brackets correspond to
 the signs of the ratios of the reduced transition matrix elements,
 $\langle f||\hat{m}(E2)||i\rangle/\langle f||\hat{m}(M1)||i\rangle$.
      Calculations are performed by using the interaction RPA2.}

\begin{center}
\begin{tabular}{||c|c||c|c||}
   \hline\hline
        Transition $i \to f$   &   $B(M1; i \to f)$ & Transition $i \to f$   &   $B(M1; i \to f)$ \\
        \hline\hline

$1^+_1\to0^+_1$ & 0.724 E--01 & $4^+_3\to3^+_1$ & 0.330 E--02 $\,\, (+)$\\

$1^+_2 \to 0^+_1$ & 0.723 E--01 & $4^+_1\to3^+_2$ & 0.299 E+00 $\,\,(-)$\\

$1^+_2\to1^+_1$ & 0.204 E+00 $\,\,(-)$ & $4^+_2\to3^+_2$ & 0.180 E+01 $\,\,(+)$\\

$2^+_1\to1^+_1$ & 0.136 E--01 $\,\,(+)$ & $4^+_3\to3^+_2$ & 0.284 E--03
$\,\,(+)$\\

$2^+_2\to1^+_1$ & 0.105 E--03 $\,\,(-)$ & $4^+_2\to4^+_1$ & 0.429 E+00
$\,\,(-)$\\

$2^+_1\to1^+_2$ & 0.346 E+00 $\,\,(-)$ & $4^+_3\to4^+_1$ & 0.272 E--03 $\,\,(+)$\\

$2^+_2\to1^+_2$ & 0.895 E--03 $\,\,(-)$ & $5^+_1\to4^+_1$ & 0.254 E+00
$\,\,(-)$\\

$2^+_2\to2^+_1$ & 0.353 E--01 $\,\,(-)$ & $5^+_2\to4^+_1$ & 0.957 E--01
$\,\,(-)$\\

$3^+_1\to2^+_1$ & 0.513 E--02 $\,\,(-)$ & $4^+_3\to4^+_2$ & 0.201 E--03
$\,\,(+)$\\

$3^+_2\to2^+_1$ & 0.433 E--01 $\,\,(-)$ & $5^+_1\to4^+_2$ & 0.121 E+01
$\,\,(+)$\\

$3^+_1\to2^+_2$ & 0.469 E--01 $\,\,(+)$ & $5^+_2\to4^+_2$ & 0.264 E+00
$\,\,(+)$\\

$3^+_2\to2^+_2$ & 0.206 E+01 $\,\,(+)$ & $5^+_1\to4^+_3$ & 0.556 E--02
$\,\,(-)$\\

$3^+_2 \to 3^+_1$ & 0.388 E+00 $\,\,(+)$ & $5^+_2\to4^+_3$ &  0.291 E--01 $\,\,(-)$\\

$4^+_1 \to 3^+_1$ & 0.184 E--03 $\,\,(-)$ & $5^+_2\to5^+_1$ &  0.250 E--01
$\,\,(+)$ \\

$4^+_2 \to 3_1$ & 0.449 E+00 $\,\,(+)$ & &\\
        \hline\hline
        \end{tabular}  \end{center}
        \end{table}

  \begin{table} 

\caption{$M1$ transition rates $B(M1)$ in units of $\mu_N^2$ between
     the levels of negative parity in $^{132}$In. Signs in brackets
     correspond to the signs of the ratios of the reduced transition matrix
     elements, $\langle f||\hat{m}(E2)||i\rangle/\langle f||\hat{m}(M1)||i\rangle$.
     Calculations are performed with the interaction RPA2.}

 \begin{center}
 \begin{tabular}{||c|c||c|c||}

            \hline\hline

        Transition $i \to f$ &  $B(M1; i \to f)$ & Transition $i \to f$ &  $B(M1; i \to f)$\\
        \hline\hline

      $2^-_1 \to 1^-_1$   &    0.384 E+01 $\,\, (+)$ & $5^-_2 \to 4^-_2$   &    0.204 E+01 $\,\, (+)$\\

      $3^-_1 \to 2^-_1$   &    0.482 E+01 $\,\, (+)$ & $5^-_2 \to 5^-_1$   &    0.795 E--02 $\,\, (+)$\\

      $3^-_2 \to 2^-_1$   &    0.166 E+00 $\,\, (+)$ & $6^-_1 \to 5^-_1$   &    0.351 E+01 $\,\, (+)$\\

      $3^-_2 \to 3^-_1$   &    0.131 E+00 $\,\, (+)$ & $6^-_2 \to 5^-_1$   &    0.223 E--01 $\,\, (+)$\\

      $4^-_1 \to 3^-_1$   &    0.470 E+01 $\,\, (+)$ & $6^-_1 \to 5^-_2$   &    0.842 E--05 $\,\, (+)$\\

      $4^-_2 \to 3^-_1$   &    0.881 E--02 $\,\, (+)$ & $6^-_2 \to 5^-_2$   &    0.956 E+00 $\,\, (+)$\\

      $4^-_1 \to 3^-_2$   &    0.575 E--02 $\,\, (-)$ & $6^-_2 \to 6^-_1$   &    0.188 E--02 $\,\, (+)$\\

      $4^-_2 \to 3^-_2$   &    0.130 E+01 $\,\, (+)$ & $7^-_1 \to 6^-_1$   &    0.257 E+01 $\,\, (+)$\\

      $4^-_2 \to 4^-_1$   &    0.189 E--01 $\,\, (+)$ & $7^-_1 \to 6^-_2$   &    0.821 E--01 $\,\, (-)$\\

      $5^-_1 \to 4^-_1$   &    0.432 E+01 $\,\, (+)$ & $8^-_1 \to 7^-_1$   &    0.134 E+01 $\,\, (+)$\\

      $5^-_2 \to 4^-_1$   &    0.482 E--04 $\,\, (-)$ & $9^-_1 \to 8^-_1$   &    0.967 E--02 $\,\, (+)$\\

      $5^-_1 \to 4^-_2$   &    0.227 E--01 $\,\, (-)$ &&\\
                  \hline\hline

  \end{tabular} \end{center}
  \end{table}

\clearpage


\end{document}